\documentclass[english,aps,prapplied,reprint,superscriptaddress,longbibliography]{revtex4-2}
\usepackage{amssymb}
\usepackage{amsmath}
\usepackage{graphicx}
\usepackage{epstopdf}
\usepackage{color}
\usepackage{braket}
\usepackage{physics}
\usepackage{mathrsfs}
\usepackage{upgreek}
\usepackage{todonotes}
\usepackage[colorlinks=true, allcolors=blue]{hyperref}
\usepackage{soul}
\usepackage{booktabs}
\usepackage{multirow}
\usepackage{float}

\DeclareMathOperator{\sinc}{sinc}
\DeclareMathOperator{\rect}{rect}

\def\thefootnote{**}\footnotetext{These authors contributed equally to this work.}

\begin{document}

\title{Demonstrating RFSoC-based Control of a Fully Connected Exchange-Only Qubit}

\author{Abigail L. Wessels\thefootnote{}}
\email[Corresponding author: ]{alwessels@hrl.com}
\affiliation{HRL Laboratories, LLC, 3011 Malibu Canyon Road, Malibu, California 90265, USA}

\author{M. Garrett Bimstefer\thefootnote{}}
\affiliation{HRL Laboratories, LLC, 3011 Malibu Canyon Road, Malibu, California 90265, USA}

\author{Joseph D. Broz}
\affiliation{HRL Laboratories, LLC, 3011 Malibu Canyon Road, Malibu, California 90265, USA}

\author{Alen Senanian}
\affiliation{HRL Laboratories, LLC, 3011 Malibu Canyon Road, Malibu, California 90265, USA}

\author{Andrew E. Oriani}
\altaffiliation{Present address: AWS Center for Quantum Computing, Pasadena, CA 91125, USA}
\affiliation{HRL Laboratories, LLC, 3011 Malibu Canyon Road, Malibu, California 90265, USA}

\begin{abstract}

	Semiconductor triple-quantum dots (TQDs) are a promising platform for scalable quantum information processing, but the community has so far lacked access to affordable and flexible high-speed control electronics. SpinQICK~\cite{spinQICK2025}, an open-source spin qubit-control package built on the Xilinx Radio Frequency System-on-Chip (RFSoC), directly addresses this gap by leveraging the high sampling-rate signal generation and advanced digital signal processing (DSP) capabilities of the Quantum Instrumentation Control Kit (QICK)~\cite{Ding2024} to enable precise time-domain voltage control and readout of semiconductor spin qubits. In this paper, we use spinQICK to demonstrate coherent control, state preparation, and readout of a fully connected triangular exchange‑only device~\cite{Acuna2024,broz2025,broz2026leakage}. We also introduce and experimentally verify a novel dynamical decoupling (DD) sequence enabled by the triangular TQD and leverage its unique decoupling properties to characterize electrical noise. These results establish spinQICK as a low-cost, flexible platform for semiconductor spin qubit control and provide a pathway for hardware optimization through qubit-level noise diagnostics.

\end{abstract}

\maketitle

\section{Introduction}\label{sec:intro}

A key challenge within quantum computing research is the lack of affordable and scalable classical control electronics. Across many qubit platforms, there is a growing need for control systems capable of fast qubit manipulation, readout and low‑latency conditional logic for feedback.  General-purpose commercial control systems are expensive and often contain proprietary elements, limiting user customization and creating a significant financial barrier for academic labs.

One path for low-cost and scalable quantum control hardware is the Xilinx Radio Frequency System-on-Chip (RFSoC) architecture.  RFSoCs couple a high-performance field-programmable gate array (FPGA) and CPU with a low-noise, high-bandwidth analog frontend via heterogeneous integration, reducing development overhead and lowering hardware costs to approximately $\mathdollar1000$ per high-speed DAC channel. Implementing a custom processor on the RFSoC enables the execution of timed instructions to orchestrate high-speed, tightly synchronized qubit manipulations and readout.  Open-source packages such as the Quantum Instrumentation Control Kit (QICK) and Qubit Control (QubiC) have leveraged this flexible and highly optimized hardware to enable control of superconducting qubits, NV-centers, and quantum dots~\cite{Stefanazzi2022, Ding2024, Xu2023, vora2024, sadi2026, koong2025}. Despite the complexity and high cost associated with control electronics, open-source qubit control solutions have not yet been widely adopted by the silicon spin qubit community.

To address the specific control requirements of gate-defined semiconductor spin qubits, HRL Laboratories has developed spinQICK~\cite{spinQICK2025}, an open-source package that uses the QICK Python API and customized firmware developed by the QICK team. This control platform is capable of charge state discrimination, spin-to-charge conversion and readout, and coherent control via both single-spin rotations and exchange operations. These features have been extended to the control of triple-quantum dot (TQD) exchange-only (EO) spin qubits, which have inherent protection from global field fluctuations, simplified baseband control, and CMOS-compatible fabrication, making them a promising platform for scalable quantum computing~\cite{Kempe2000,DiVincenzo2000,FongWadzura2011,Weinstein2023,Madzik2025,hrl2026digitally}.

In this paper, we demonstrate the capabilities and flexibility of spinQICK and the RFSoC platform in the coherent control of a fully connected triangular TQD device~\cite{Acuna2024}. These devices, which allow all-to-all coupling and three axes of exchange, have potential advantages such as leakage protection, shorter gate operations, and denser packing when compared to linear TQD devices~\cite{Weinstein2005,broz2025,Chadwick2025,broz2026leakage}. Triangular devices also offer unique control challenges, such as enhanced cross-coupling between neighboring dots, making them an ideal testbed for the capabilities of the RFSoC.


Demonstrating and validating the capabilities of the RFSoC platform requires meaningful metrics of both low‑frequency and high‑frequency electrical noise. Dynamical decoupling (DD) sequences provide a way to extend coherence times in an EO qubit, and also act as a test of pulse axis alignment, timing and calibration. However, conventional finite‑pulse DD sequences on linear TQDs, such as NZ1y~\cite{Sun2024}, suppress the effects of the instrumentation-induced gate-referred voltage noise we seek to characterize. The triangular device allows implementation of the NZM1 sequence, introduced here, which fully decouples hyperfine noise while retaining a low-frequency charge noise error channel. This makes NZM1 a sensitive probe for characterization of voltage noise from control electronics.

\begin{figure*}[t!]
	\includegraphics[width=1.9\columnwidth]{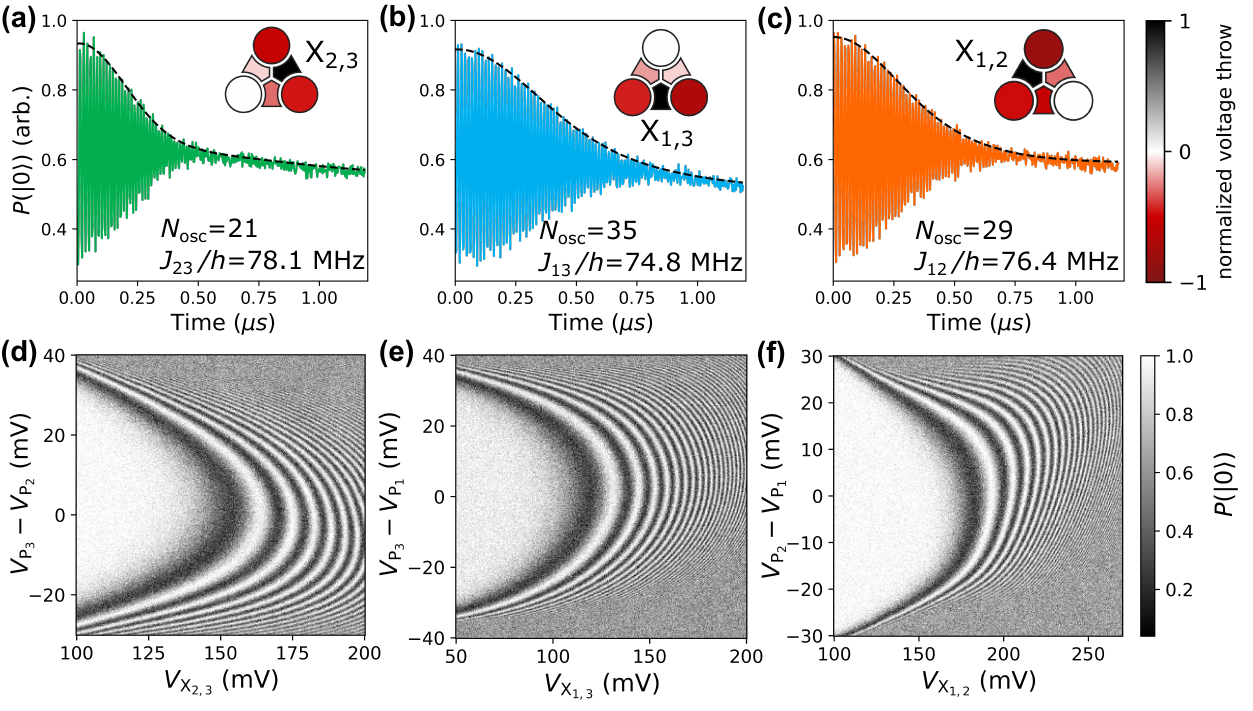}
	\caption{{\bf SpinQICK EO Qubit Tune-Up:} (a-c) Coherent exchange oscillations due to $J_{23}$, $J_{13}$, and $J_{12}$ exchange couplings of a triangular device, respectively. The parameter $N_{\mathrm{osc}}$ given for each axis is defined as the number of exchange oscillations before the amplitude decays by $1/e$. Insets: diagrams showing the normalized amplitude of pulses played on each gate of the triangle device to generate a pulse along the defined symmetric exchange axis. The negative voltages are played using the crosstalk compensation feature present in firmware. (d-f) Fingerprint plots showing exchange oscillations due to the $J_{23}$, $J_{13}$, and $J_{12}$ exchange couplings by sweeping exchange pulse amplitude and double-dot detuning. Exchange pulse duration was $10~\mathrm{ns}$ for each plot.}
	\label{fig_1}
\end{figure*}

Using NZ1 and NZM1 together, we perform initialization‑dependent DD measurements to characterize hyperfine and charge noise, which enables quantification of both high‑frequency white noise and drift in the measurement setup. These measurements are performed using modified QICK firmware optimized for spin qubit control in conjunction with low-cost commercially available off-the-shelf (COTS) analog electronics to amplify and condition signals from the RFSoC. Measurements of initialization-dependent decoherence are compared with stochastic open quantum system simulations using noise spectra extracted from qubit characterization experiments. We find agreement between simulation and experiment and in doing so we demonstrate the capabilities and flexibility of the RFSoC and the fully open-source spinQICK platform.

\section{spinQICK Control Platform}\label{sec:spinQICK}

We implement a custom software stack, coupled with firmware developed by the QICK team~\cite{Stefanazzi2022,Ding2024}, to enable state preparation, readout, and coherent control of spin qubits using a Xilinx ZCU216 RFSoC evaluation board.  This stack, collectively known as spinQICK, features 16 high-speed digital-to-analog converters (DACs) with sampling rates of $6.88~\mathrm{GSPS}$. Of these channels, 14 are optimized for baseband waveform generation, and two channels feature direct digital synthesis (DDS) capability for generating the readout stimulus. The firmware used for this demonstration also features a DSP feature which enables crosstalk compensation on 11 of the baseband pulsing channels. SpinQICK consists of a library of control experiments that utilize the QICK processor which runs on the FPGA and executes timed instructions from a user-generated program. This enables users to program a library of pulses and readout commands that can be played at specified relative times on a $2.5~\mathrm{ns}$ clock. The processor can also perform loops, branching, conditional logic, and arithmetic operations.  The high-speed differential DACs and analog-to-digital converters (ADCs) are connected to our experimental setup with COTS DC-coupled amplifiers as described in App.~\ref{sec:expmethods:setup}.

Readout is done via one of the 11 available analog-to-digital converters (ADCs), which sample at $2.5~\mathrm{GSPS}$. Charge state readout is done by measuring the conductance of a dot-charge-sensor (DCS) (see App.~\ref{sec:expmethods:setup} for a diagram of the readout circuitry). To perform this conductance measurement with spinQICK, the RFSoC generates a $1~\mathrm{MHz}$ AC signal using a DAC channel with DDS capability, and digitizes the return signal with an ADC. It then demodulates and integrates the signal over the full length of the readout pulse using DSP in the ADC data pipeline. This greatly reduces the amount of data that needs to be streamed from the board and handled by software after data acquisition.

Spin-to-charge conversion is performed using Pauli spin blockade readout as described in Ref.~\cite{Blumoff2022}. We use the RFSoC's high-speed DACs to pulse to different points in charge stability space, as described in App.~\ref{sec:expmethods:spintocharge}. Using this method to carry out state preparation and measurement, we obtained typical signal-to-noise ratios of 5 to 6 in this work.

Coherent control of EO qubits is performed by modulating the exchange interaction between neighboring electrons using baseband voltage pulses~\cite{Weinstein2023}. To perform exchange pulses on our qubit, we play predistorted~\cite{Guo2024} $10~\mathrm{ns}$ baseband pulse waveforms pre-programmed into QICK's waveform memory. We used the crosstalk compensation DSP feature in our firmware, which allows for automatic linear compensation between gates based on a user-definable cross-coupling matrix, to define the symmetric exchange axis~\cite{Reed2016}. This allowed us to virtualize the exchange gates~\cite{hensgens2017quantum} and to apply the compensating pulses on neighboring gates without explicitly including these pulses in our QICK program. To mitigate axis misalignment problems we implemented compensation between exchange gates as well \cite{broz2025}, as described in App.~\ref{sec:expmethods:exchangecal}. Panels (a-c) of Fig.~\ref{fig_1} show exchange oscillations for each axis, and insets illustrate the amplitude of the compensating pulses relative to the pulse on the exchange gate. Panels (d-f) of Fig.~\ref{fig_1} show the resulting fingerprints for each of the three exchange axes that were taken with spinQICK after the cross-coupling matrix was set. After defining each exchange axis, exchange angle calibrations were performed with the methods described in Ref.~\cite{Andrews2019}. With all three axes calibrated, we were able to perform the novel DD demonstrations described in this work.

In the following sections, we describe the operation of triangular EO qubits and DD pulse sequences that can be performed on them. We will apply these DD sequences using spinQICK and RFSoC control and readout to study the effects of device and system-level noise.

\section{Triangular EO Qubit}\label{sec:EOqubit}

\begin{figure}[t]
	\includegraphics[width=\columnwidth]{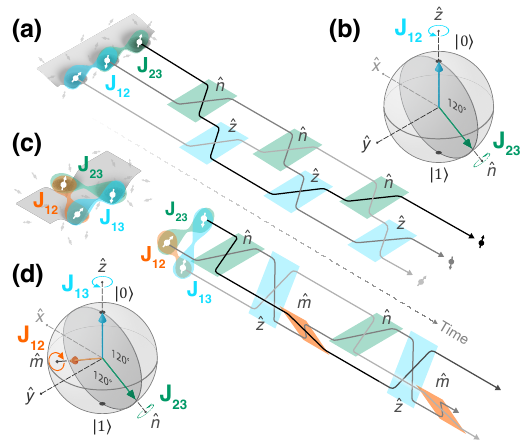}
	\caption{{\bf spinQICK DD Implementation:} (a) Conventional EO qubits have a linear arrangement and control is implemented with two exchange couplings, $J_{12}$ and $J_{23}$. These two couplings can be used to make a DD sequence which decouples from quasistatic exchange noise. A triangular EO qubit gives an additional exchange coupling, $J_{13}$, and can be used to create a DD sequence which decouples from quasistatic hyperfine noise, as shown in (c). White and grey arrows represent the EO qubit and hyperfine noise spins, respectively. In the Bloch sphere representation, the linear geometry (b) enables rotations about the $\hat{z}$ and $\hat{n}$ axes whereas the triangular geometry (d) provides a third $\hat{m}$ axis.}
	\label{fig_2}
\end{figure}

Much of the experimental focus in EO qubits has been on linear layouts with only two exchange couplings, providing the two non-orthogonal rotation axes $(\hat{z},\hat{n})$ required for universal control, as shown in Fig.~\ref{fig_2}(a,b)~\cite{Eng2015,Andrews2019,Weinstein2023}. However, recent experimental developments have shown coherent control of an EO qubit in a fully connected triangular TQD~\cite{Acuna2024}, as depicted in Fig.~\ref{fig_2}(c,d). While this high-connectivity layout is not necessary for universal control, it allows many possibilities for novel EO qubit operational methods such as always-on exchange and geometric encodings~\cite{Scarola2004,Weinstein2005,Hsieh2012,broz2025}.

To show the unique properties of triangular TQDs, we first generalize the linear EO qubit Hamiltonian to include the additional connectivity of the triangular TQD. Panels (a,c) of Fig.~\ref{fig_2} show qualitative depictions of the TQD potential energy landscapes which confine and control the three spins that form an encoded EO qubit. Assuming one electron per dot, these TQD systems can be generally expressed as a Heisenberg spin lattice,
\begin{equation}\label{eq:H_0}
	H_{\mathrm{0}} = \sum_{i,j}J_{ij}({\bf S}_{i}\cdot{\bf S}_{j})-\gamma{\bf B}\cdot\sum_{i}{\bf S}_{i},
\end{equation}
\noindent where $i$ indexes each dot and $(i,j)$ indexes exchange-coupled dots. ${\bf S}={\boldsymbol\sigma}/2$ is the spin operator, $J_{ij}$ is the exchange coupling, $\gamma$ is the gyromagnetic ratio, and ${\bf B}$ is an external magnetic field. In this report, we use ${\bf B} = B_{z} = 50~\mathrm{\mu T}$ to represent Earth's magnetic field. For the triangular TQD, we arrive at the Hamiltonian,
\begin{align}\label{eq:H_tri}
	H_{\mathrm{\Delta}} & = [J_{12}({\bf S}_{1}\cdot{\bf S}_{2})+J_{23}({\bf S}_{2}\cdot{\bf S}_{3}) \nonumber \\
	                    & +J_{13}({\bf S}_{1}\cdot{\bf S}_{3})]-\gamma B_{z}S_{z},
\end{align}
\noindent where, in general, there is an additional term in Eq.~\ref{eq:H_tri} accounting for the Aharonov-Bohm phase from enclosed magnetic flux~\cite{Scarola2004}. However, this term is negligible if the magnetic field vector is in the plane of the TQD array.

The three-spin basis for EO qubits can be conveniently expressed in terms of spin quantum numbers $\ket{S_{12},S,m_z}$. The total spin $S$ of the three-spin system can take values of $1/2$ or $3/2$, with the four $S=1/2$ states comprising a decoherence-free subsystem (DFS); the remaining four leakage states with $S=3/2$ are not intentionally used. The DFS  states are specified by the total spin of the first two spins $S_{12}=(0,1)$ and the total $z$-spin $m_z=\pm 1/2$, with $S_{12}$ giving the encoded qubit state and $m_z$ the gauge qubit state. The 4-dimensional DFS is the tensor product of two two-level systems.  Pauli spin blockade readout determines the value of $S_{12}$. Because exchange commutes with all global interactions, and in particular global exchange and global magnetic interactions, global interaction Hamiltonians must be proportional to the identity in each $S$ block, thereby preserving all encoded information. $m_z$ does generally evolve under global interactions, but the subsystem structure guarantees that encoded information does not evolve or become entangled with the gauge under global interactions.

Modulating the exchange coupling between a spin pair induces controlled singlet-triplet oscillations, corresponding to rotations in the DFS-encoded Bloch sphere representation. The rotation axis depends on the interacting spin pair, namely $(1,3)\rightarrow\hat{z}$, $(2,3)\rightarrow\hat{n}=-(\sqrt{3}\hat{x}+\hat{z})/2$, and $(1,2)\rightarrow\hat{m}=(\sqrt{3}\hat{x}-\hat{z})/2$, as shown in Fig.~\ref{fig_2}(d).

Linear TQDs can only activate $J_{12}$ and $J_{23}$ whereas triangular TQDs enable $J_{13}$. While this additional exchange axis gives rise to interesting physics, the required close-packing of the dots leads to weaker electron confinement and increased cross capacitance. This means careful inter-gate compensation is required to maintain well-defined potentials to reduce orbital excitations and maintain a Heisenberg spin lattice~\cite{Acuna2024,broz2025}. This problem, which adds complexity to pulse programming, is well-suited for spinQICK and the RFSoC, which can efficiently address these non-idealities in real time via crosstalk compensation DSP capabilities discussed in Sec.~\ref{sec:spinQICK}. As a test of the spinQICK platform's EO qubit control performance, we use DD to both benchmark the platform's capabilities and investigate the qubit noise environment.

\section{Dynamical Decoupling Test}\label{sec:DD}

While the logical EO qubit is immune to global noise which affects all three physical spins equally, inhomogeneous noise can lead to relaxation and dephasing~\cite{West2012}. DD sequences in EO qubits globalize inhomogeneous noise by averaging the noise environment in each dot across all spins, analogous to Hahn echo~\cite{Hahn1950,Meiboom1958,Sun2024}. As shown in Fig.~\ref{fig_2}(a,c), this can be accomplished with a set of site-wise spin swaps (i.e., $\pi$ rotations, e.g., $R_{\hat{z}}(\pi)=\mathrm{Z}$) where each spin equally samples the local noise environment from each dot. After the sequence, the qubit returns to the initial state with each spin accumulating noise from each dot, leading to noise decoupling depending on the initial qubit state and sequence implemented.

The dominant noise sources in EO qubits are imperfect inter-spin exchange couplings and inhomogeneous magnetic fields. Magnetic noise originates from inhomogeneous hyperfine interactions between spinful nuclei in the semiconductor heterostructure and the three physical spins which constitute the logical EO qubit~\cite{Taylor2007,Andrews2019,Kerckhoff2021}. Hyperfine noise is intrinsic to the TQD device, so the control electronics should have no direct influence on its magnitude. Exchange noise, however, originates from electrical noise which varies the TQD potentials, such as control voltage and gate dielectric charge noise~\cite{Glattli1997,Andrews2019,Reed2016}. An optimized control system's noise contribution should be low enough such that the intrinsic charge noise component dominates the total exchange noise.

To investigate these noise sources, we use DD sequences to perform \textit{in-situ} noise characterization akin to qubit noise spectroscopy~\cite{Bylander2011,Kerckhoff2021,Ezzell2023,Sun2024,Vezvaee2024}. DD sequences most effectively decouple from noise which varies slowly relative to the length of the sequence, known as quasistatic noise~\cite{Merkulov2002}. Consequently, DD sequences can be represented as noise filters, typically exhibiting a high-pass or band-pass character with frequency cutoff corresponding to the sequence repetition rate~\cite{Biercuk2011}. In this report, we leverage the unique decoupling capabilities of the triangular EO qubit to perform a suite of noise investigations, allowing the characterization of both hyperfine and exchange noise using two decoupling sequence types.

The standard EO qubit DD sequence $[NZ]^{3}$, denoted NZ1, originally reported in Ref.~\cite{West2012}, is shown in Fig.~\ref{fig_2}(a). This sequence decouples from quasistatic exchange noise and is susceptible to quasistatic hyperfine noise for finite pulse times via a $\sigma_{y}$ error channel, providing full decoupling from quasistatic noise if initialized along the $\hat{y}$ axis of the Bloch sphere~\cite{Sun2024}. This allows one to interrogate low-frequency hyperfine noise and high-frequency noise by comparing the qubit decay envelopes between initializations. With a triangular device, it is possible to construct a similar DD sequence which decouples from quasistatic hyperfine noise and decays from quasistatic exchange noise through a ${\sigma_{z}}$ error channel, enabling exchange noise characterization. This novel $[NZM]^{2}$ sequence, denoted NZM1, is shown in Fig.~\ref{fig_2}(c). Conceptually, this is due to the unique ability of triangular TQDs to directly average the local magnetic field between all sites, whereas linear TQDs can only average between sites $(1,2)$ and $(2,3)$.

\begin{figure*}[t]
	\includegraphics[width=1.8\columnwidth]{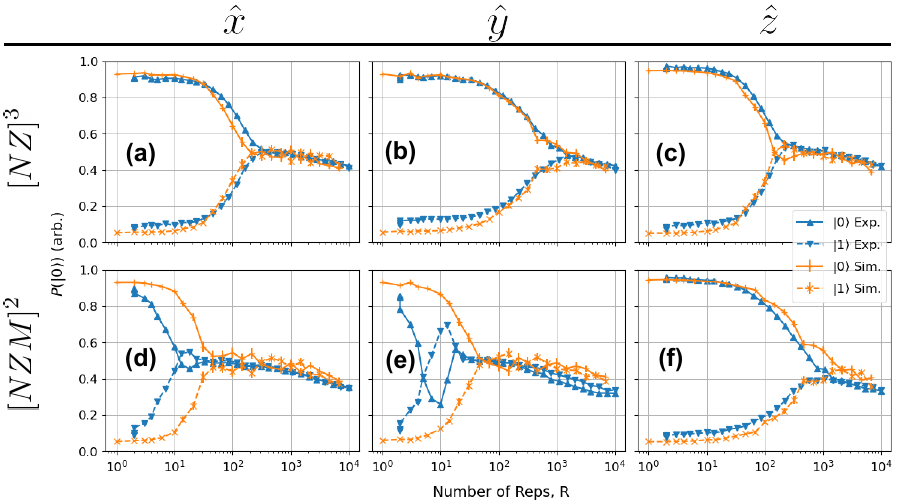}
	\caption{{\bf NZ1 and NZM1 Experiments:} Initialization-dependent decay envelopes for NZ1 and NZM1 where (a-c) and (d-f) give their respective return fidelities for initializations along the $\hat{x}$, $\hat{y}$, and $\hat{z}$ axes of the Bloch sphere. Initialization in the $\ket{1}$ state is implemented with a $\mathrm{[DD]}X$ pre-rotation, where $\mathrm{[DD]}$ represents a DD sequence. Initializations along $\hat{x}$ and $\hat{y}$ are implemented with pre- and post-rotations $H\mathrm{[DD]}H$ and $HS^{\dagger}\mathrm{[DD]}SH$, respectively.}
	\label{fig_3}
\end{figure*}

We used spinQICK to experimentally measure the initialization-dependent probability of returning the initial state, known as the return fidelity, of the NZ1 and NZM1 sequences and extract information about the qubit noise environment. App.~\ref{sec:expmethods:exchangecal} contains details pertaining to executing these experiments with QICK. We also performed computational simulations which reflect experimental conditions, as described in App.~\ref{sec:mathmethods:sim}. DD sequences were simulated stochastically with shot-to-shot uncorrelated noise trajectories sampled from noise power spectral densities (PSDs) extracted from qubit measurements, as described in App.~\ref{sec:mathmethods:noise}. The results from these simulations are given in Fig.~\ref{fig_3} along with the corresponding experimentally measured data.

\section{Discussion}\label{sec:discussion}

Starting with the NZ1 results in Fig.~\ref{fig_3}~(a-c), we observe agreement between simulation and experiment, implying spinQICK's timing and inter-gate crosstalk compensation are meeting the requirements for EO qubit operation. For NZ1 initialized along $\hat{y}$, denoted NZ1y, we observe an exponential decay envelope (e.g., $C(t)\propto e^{-at}$) and longer lifetime as opposed to the Gaussian decay envelope (e.g., $C(t)\propto e^{-at^{2}}$) seen in NZ1 initialized along $\hat{x}$ and $\hat{z}$, denoted NZ1x and NZ1z, respectively. This behavior follows directly from each sequence's sensitivity and efficacy in decoupling from hyperfine noise, with the NZ1x and NZ1z decay envelopes being sensitive to quasistatic hyperfine noise, and NZ1y decoupling from this $\sigma_{y}$ error channel and decaying only from high-frequency exchange and hyperfine noise (see App.~\ref{sec:mathmethods:noise} for details).

Interestingly, the NZ1y lifetime measured in this report ($T_{2}^{\mathrm{NZ1y}}\approx60~\mathrm{\mu s}$) is approximately an order of magnitude lower than that measured in Ref.~\cite{Sun2024} ($T_{2}^{\mathrm{NZ1y}}\approx600~\mathrm{\mu s}$, Fig. 4(b) $t_{\mathrm{idle}}=40~\mathrm{ns}$) using a device with similar intrinsic noise. We attribute this discrepancy to the fitted high-frequency white noise floor of approximately $40~\mathrm{nV/\sqrt{Hz}}$ which was not present in Ref.~\cite{Sun2024}. The room-temperature amplification chain, consisting of Texas Instruments LMH5401 and THS3491 amplifiers, likely contributes significantly to this noise floor. The chain provides an effective combined gain of 20 V/V and has an input-referred noise of $1.9~\mathrm{nV/\sqrt{Hz}}$, according to the data-sheet specifications.  These amplifiers were used to amplify RFSoC output pulses to reach exchange rates of at least $100~\mathrm{MHz}$ at the device. This highlights a need for a lower-noise pulse amplification solution to optimize the spinQICK platform.

Having investigated the high-frequency regime, we now explore the contribution of control electronics noise in the low-frequency regime using the NZM1 sequence. NZM1 initialized along $\hat{z}$, denoted NZM1z, has similar decoupling characteristics to NZ1y and yields similar insights. NZM1x and NZM1y are sensitive to quasistatic exchange noise and overrotation, so we expect to see a different lifetime indicative of the noise magnitude and, if time-correlated, coherent return fidelity oscillations. The NZM1x and NZM1y experimental data given in Fig.~\ref{fig_3}(d-e) deviate from simulation and oscillate, implying shot-to-shot correlated noise not captured by simulation. Gate-referred voltage noise in semiconductor quantum dots is typically well-described by zero-mean Gaussian random variables and characterized by a $1/f$ PSD in systems limited by gate dielectric charge noise~\cite{Glattli1997}. However, our measured voltage noise PSD fits a Lorentzian with amplitude and correlation decay time of approximately $300~\mathrm{\mu V}$ and $200~\mathrm{ms}$, as shown in Fig.~\ref{fig_A1}(b).

Correlating the magnitude of overrotation over time, as described in App.~\ref{sec:mathmethods:drift}, we find an approximately linear trend over long timescales with a slope of $65.7~\mathrm{Hz/s}$ and y-intercept of $-1.7~\mathrm{kHz}$, implying a good initial calibration which drifted and led to the observed return fidelity oscillations. We note that the same calibrated $\pi$-pulse voltages were used for the duration of data collection shown in Fig.~\ref{fig_3}. Deconvolving the source of this noise is difficult due to the use of a high-speed adder, identical to that used in Ref.~\cite{Blumoff2022}, as it could be due to drift in the precision DC sources or a differential voltage between the high- and low-speed sides of the adder from non-optimal grounding. The device and signal chain were the same as those used in Ref.~\cite{broz2025}, except that the pulsing and readout instrumentation was replaced by the RFSoC and LMH5401 differential amplifiers. The $N_{\mathrm{osc}}$ values given in Ref.~\cite{broz2025} are similar to those presented in this work, implying similar short-timescale performance. However, the randomized benchmarking results given in Ref.~\cite{broz2025} cannot identify drift since overrotation only reduces the observed gate fidelity and is consequently indistinguishable from incoherent noise~\cite{Andrews2019}. Therefore, the drift we observed may originate from the room-temperature amplification chain, which will be resolved in future system design revisions.

\section{Conclusion}\label{sec:conclusion}

In summary, we have demonstrated that the spinQICK control platform, built on the Xilinx RFSoC, is a promising high-performance, low-cost, and accessible alternative for control of gate-defined spin qubits. We leveraged the unique properties of triangular EO qubits to perform a suite of DD measurements to separately quantify hyperfine and exchange noise, enabling direct identification of high‑frequency noise originating in the amplification chain as well as slow drift associated with low‑frequency electrical fluctuations. Comparison with stochastic simulations using experimentally extracted noise spectra further validates the accuracy of the control stack and the reliability of our noise‑characterization methods.

Although long‑timescale stability remains an avenue for improvement, the results presented here establish spinQICK as an accessible and effective control solution for semiconductor spin qubits. The ability to diagnose noise sources and system drift directly through qubit‑level measurements provides a powerful mechanism for guiding future hardware optimization. Overall, this work advances RFSoC‑based quantum‑control technology and highlights the role of open‑source instrumentation in enabling scalable and precise spin‑qubit experiments.

\vspace{-1em}
\section*{Acknowledgements}
The sample used in this experiment was made by the HRL device fabrication team as part of the Qubits for Computing Foundry program. The authors thank John Carpenter for assisting with the preparation of figures, Bryan Fong and Tyler Keating for advising on the DD work, Thaddeus Ladd for reviewing the paper's content and providing feedback, and Jason Petta for his early feedback and guidance on qubit control, discussions of triangle device physics, and for reviewing this paper and providing feedback. We also thank Sho Uemura, Gustavo Cancelo, Kenneth Treptow, Diego Martin, and Luis Arnaldi for their continued support of custom QICK firmware. We also thank Edwin Acuna, Aaron Chronister, and Daniel Sanchez for valuable discussions on the topics presented in this work.

\appendix

\vspace{-1em}
\section{Experimental Methods}\label{sec:expmethods}
\subsection{Experimental Setup}\label{sec:expmethods:setup}
\vspace{-0.5em}
\renewcommand\thefigure{\thesection\arabic{figure}}
\setcounter{figure}{0}

In this work, we used an AMD Xilinx ZCU216 evaluation board to generate all time-varying voltage signals applied to the device. This board features a Zynq Ultrascale+ RFSoC capable of generating waveforms at sampling rates up to 9.85 GSPS on 16 output channels, and digitizing signals on 16 input channels at 2.5 GSPS. To coordinate and execute high-speed pulses and readout we use custom firmware developed for this RFSoC by the QICK team~\cite{Ding2024}. The timed processor used by QICK has special registers for implementing arithmetic operations, random number generation, and conditional logic operations and branching, and features a 256 depth loop counter for efficient instruction usage. Instructions are compiled into machine code via the QICK Python API, and executed at a rate of $430~\mathrm{MHz}$. For spin qubit control, this processor is supplemented with modified DSP blocks which enable 14 high-speed (6.8 GSPS) DAC channels without DDS to maximize wave memory for baseband envelope generation, and 2 DAC channels with DDS enabled for the AC readout signal. Crosstalk compensation is implemented in DSP for 11 of the 16 baseband pulsing channels, and all off-diagonal elements in the $11\times11$ crosstalk matrix are user configurable. This firmware is available for download at Ref.~\cite{spinQICK2025}.

The ZCU216 DACs generate high-speed signals in the form of differential pair outputs with a voltage range of approximately $800~\mathrm{mV}$ peak-to-peak. A Texas Instruments LMH5401 amplifier evaluation module is used to

\begin{figure}[t!]
	\includegraphics[width=0.85\columnwidth]{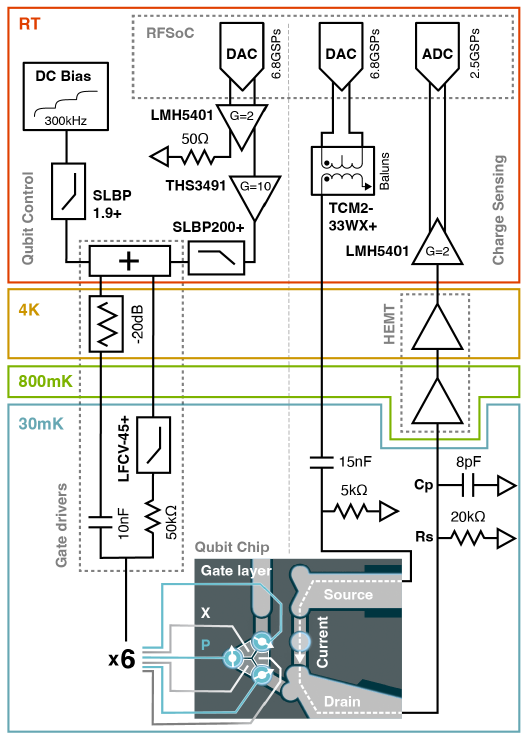}
	\caption{{\bf spinQICK EO Measurement Setup:} A complete diagram of the room-temperature and cryogenic control hardware used in this paper. Baseband pulses from the RFSoC are added via a circuit composed of a high-bandwidth voltage divider at room temperature and a resistive bias tee at mK. A more detailed explanation of this ``adder" circuit can be found in Ref.~\cite{Blumoff2022}. Charge sensing is done by passing an AC current through the dot-charge sensor, with the stimulus provided by an RFSoC DAC through a $10~\mathrm{MHz}-1~\mathrm{GHz}$ balun. Current-to-voltage conversion is done via a $20~\mathrm{k\Omega}$ resistor to ground. The approximately $8~\mathrm{pF}$ parasitic capacitance to ground sets the cutoff of the stimulus signal at approximately $1~\mathrm{MHz}$. Readout is done utilizing an RFSoC ADC, with demodulation done in real time through digital downconversion and averaging. A more detailed explanation of this digital signal chain can be found in Ref.~\cite{Stefanazzi2022}. DC bias is provided through a custom 20-bit DC source.}
	\label{Meas_Setup}
\end{figure}

\noindent convert the differential pair to a DC-coupled single-ended output. This module is configured to have a gain of $G=2$ when using a single output and provides a matched $50~\Omega$ output. We use the Texas Instruments THS3491 amplifier module with a gain configuration of $G=10$ as a second stage amplifier to counter the voltage division designed into our existing high-speed adder circuit design~\cite{Blumoff2022}. A diagram of the experimental setup is shown in Fig.~\ref{Meas_Setup}.

\vspace{-1em}
\subsection{Readout and Spin to Charge Conversion}\label{sec:expmethods:spintocharge}
\vspace{-0.5em}

To perform spin-to-charge conversion, we used standard Pauli spin blockade (PSB) techniques and discriminated the double-dot charge state with a dot-charge sensor (DCS) as described in Ref.~\cite{Blumoff2022}. We do this by applying a small $1~\mathrm{MHz}$ pulse generated by the RFSoC to the source side of the DCS. The current through the DCS is converted to a voltage via a $20~\mathrm{k\Omega}$ resistor to ground on the drain side of the DCS and amplified both cryogenically and at room temperature before being digitized by an RFSoC ADC channel. The digitized signal is demodulated on the RFSoC into magnitude and phase quadratures and integrated over the duration of the pulse. For our measurements we take the quadrature sum of these components to get an effective conductance measurement. This demodulation occurs in real time, greatly reducing latency and the need for data transfer and off-board handling.

\begin{figure}[t!]
	\includegraphics[width=\columnwidth]{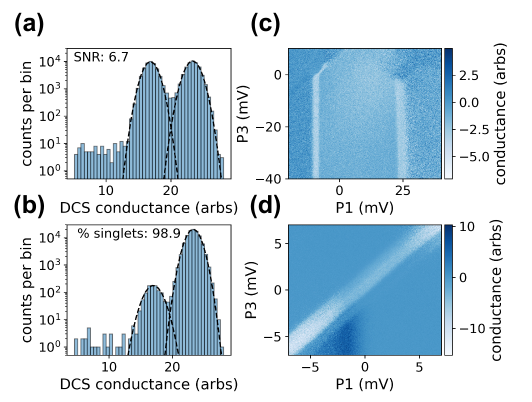}
	\caption{{\bf Pauli spin blockade readout:} (a) A histogram of $10^{5}$ measurements after preparing a singlet, going to the (1,1) charge cell and dephasing for $10~\mathrm{\mu s}$. The signal-to-noise ratio is 6.7. (b) Histogram of $10^{5}$ measurements after preparing a singlet and measuring at the measurement window. The percentage of counts categorized as singlets is 98.9$\%$. (c) 2D sweep of the singlet initialization point in the spam sequence. The color represents the difference in conductance between an initialized state and the state after dephasing in the (1,1) charge configuration. (d) 2D sweep of the measurement point, performed with the same sequence of steps.}
	\label{fig_psb}
\end{figure}

SpinQICK makes use of real-only generators that remove direct-digital synthesis and phase control capabilities to free up FPGA resources for DSP. These baseband pulses are used to diabatically condition the double-dot charge state on the two-electron spin state using PSB. All PSB measurements were performed relative to a reference measurement immediately preceding each shot of the experiment to mitigate low-frequency noise. This reference measurement consists of initializing a pair of electrons in a singlet state and then measuring at the $(2,0)-(1,1)$ charge boundary, referred to as the measurement window. We threshold the differential conductance measurement to assign the measurement as a singlet or triplet as shown in Fig.~\ref{fig_psb}(a). We were able to consistently obtain signal-to-noise ratios of 5 to 6 in our measurements in this work, with a typical measurement histogram shown in Fig.~\ref{fig_psb}(a) and a state preparation measurement shown in (b). We perform a sweep of the initialization point in this sequence to obtain an optimal position for fast initialization as shown in Fig.~\ref{fig_psb}(c). Likewise, to identify the voltage coordinates of the measurement window, we perform a sweep of the measurement point as shown in Fig.~\ref{fig_psb}(d).

To overcome limitations associated with QICK's limited per-channel waveform memory, spinQICK updates the amplitude of a minimum-length pulse ($7.5~\mathrm{ns}$) stored in waveform memory, and the amplitude remains at that value until the next operation is executed.  This approach allows users to play pulses that exceed the approximately $10~\mathrm{\mu s}$ of per-channel memory QICK allocates by default. Sequences which include long ramps or sweeps of ramp values can be implemented via a series of steps. The measurement window sweep shown in Fig. \ref{fig_psb}(d) is an example of such a sweep, where ramps into and out of the readout window are used to avoid Landau-Zener transitions into unwanted states.

\begin{figure}[h!]
	\includegraphics[width=0.8\columnwidth]{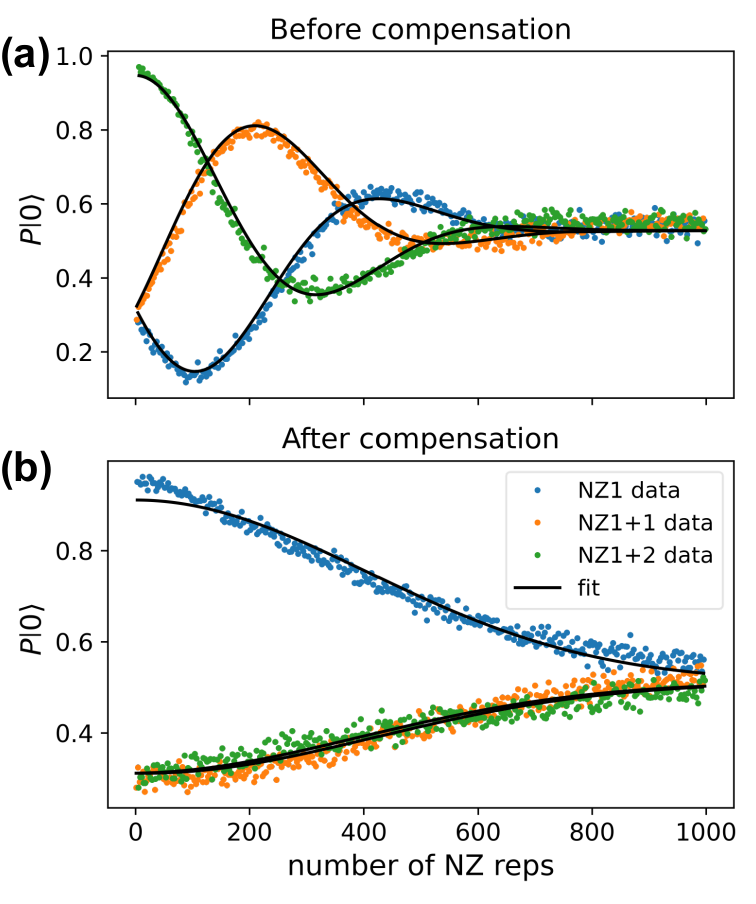}
	\caption{{\bf Measuring axis misalignment:} Singlet return probability vs number of $NZ$ reps. The data is broken up into NZ1 (blue points), NZ1+1 (orange points), and NZ1+2 (green points) branches. (a) The result without compensating on neighboring x gates. The fit to Eq.~\ref{eq:misalign} is shown, with a fitted axis misalignment of $0.24^{\circ}$. (b) The same sequence, after adding compensation on neighboring x gates and calibrating both axes again.}
	\label{fix_axis_mis}
\end{figure}

\vspace{-1em}
\subsection{Exchange pulses with spinQICK}\label{sec:expmethods:exchangecal}

As mentioned in Sec.~\ref{sec:spinQICK}, instead of programming and playing pulses on neighboring gates while pulsing the exchange gates ($\mathrm{X_{1,3}}$, $\mathrm{X_{1,2}}$, $\mathrm{X_{2,3}}$), we used the crosstalk compensation feature built into the QICK firmware to define the symmetric exchange axis. This decreased the complexity of the program that needed to be written to perform the DD experiment. We found that the triangle device was particularly susceptible to axis misalignment effects. We identified this by plotting all three ``branches'' of the NZ1 experiment as a function of $NZ$ repetitions, as shown in Fig.~\ref{fix_axis_mis}. The oscillations in these data are well described by the following equation, which describes an $NZ$ sequence in the presence of an axis misalignment

\begin{equation}\label{eq:misalign}
	P_{\ket{0}} = a_0+a\cos\left[ x \left(\frac{2\pi}{3} + 2 \delta\theta \right) \right] \exp\left[\frac{-x^2}{\tau^2}\right]
\end{equation},

where $x$ is the number of $NZ$ repetitions, $\delta\theta$ is the axis misalignment in radians, $\tau$ is the decay constant, $a_0$ is the asymptotic value, and $a$ is the amount of contrast at $x=0$. To combat this misalignment, we implemented compensation between exchange gates on this device in addition to the compensation on neighboring plunger gates, as described in~\cite{broz2025}. We tuned this by increasing the cross-coupling matrix coefficients until we no longer observed oscillations in the NZ1 sequence data. We repeated the process using the M axis in place of the $N$ axis to obtain the matrix elements for the third axis.

\begin{figure}[t!]
	\includegraphics[width=0.75\columnwidth]{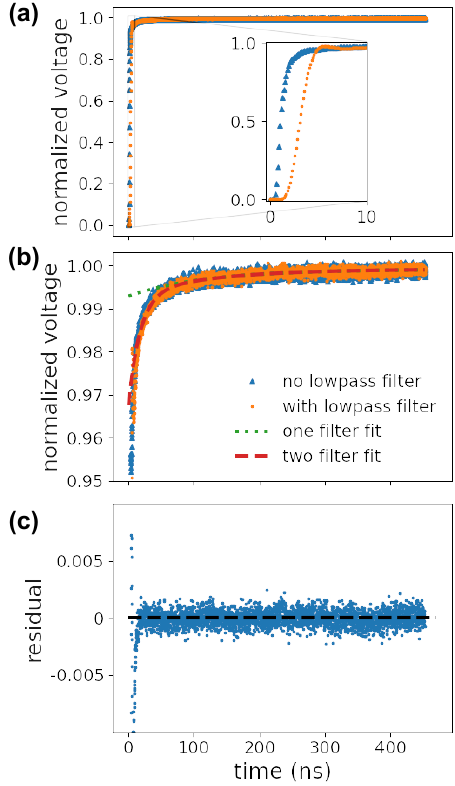}
	\caption{{\bf Room-temperature pulse characterization:} (a) Normalized $500 ~\mathrm{ns}$ time trace of a baseband pulse generated by the RFSoC after being amplified by the LMH5401 and THS3491 amplifiers with (orange circles) and without (blue triangles) the SLBP 200+ low-pass filter in the signal chain. (b) The same data, zoomed in to show the fit used to calculate the IIR filter parameters. The green dotted line shows the result with a fit to the $ >100~\mathrm{ns}$ data to capture the longer time constant, and the red dashed line shows a fit including this time constant and a second pole capturing the $ >10~\mathrm{ns}$ behavior. (c) The residual of the two-pole fit.}
	\label{fig_RTP_char}
\end{figure}

\begin{figure}[h!]
	\includegraphics[width=0.75\columnwidth]{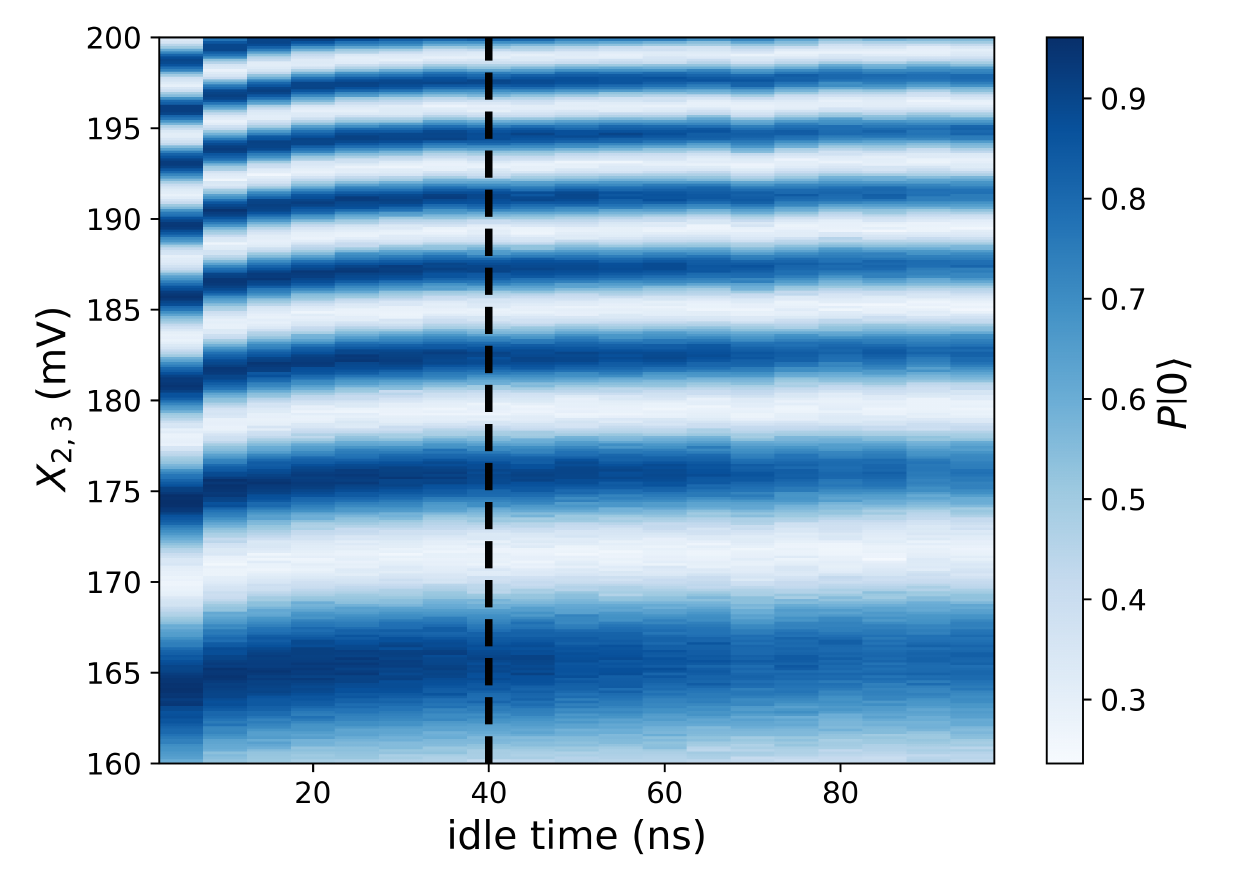}
	\caption{{\bf Cryogenic pulse characterization:} Singlet return probability after a series of 10 $10 ~\mathrm{ns}$ voltage pulses as a function of exchange gate voltage and time between pulses. The dashed line indicates the $40 ~\mathrm{ns}$ idle time which was chosen for the DD demonstrations in this work.}
	\label{fig_Cryo_char}
\end{figure}

To implement exchange pulses with the spinQICK control platform, we programmed $10~\mathrm{ns}$ baseband pulse shapes fully into QICK's waveform memory. The ability to store full pulses and short pulse sequences allowed us to apply pulse predistortion to the waveforms to reduce the effects of signal distortion on timescales of the pulse length or longer. To obtain the filter we applied to predistort pulses, we used a sampling oscilloscope to collect pulse-shape data from the room-temperature signal chain consisting of the ZCU216 \textrightarrow LMH5401 \textrightarrow THS3491 \textrightarrow SLBP 200+ low-pass filter as shown in Fig.~\ref{Meas_Setup}. This data is shown in Fig. \ref{fig_RTP_char}. We then fit this data using the method described in Ref.~\cite{Guo2024} and obtained coefficients for two single-pole infinite impulse response (IIR) filters. To determine whether we had any significant interference between pulses applied in series, we performed a sweep of exchange pulse amplitude for a set of 10 pulses while sweeping the idle time between pulses as shown in Fig.~\ref{fig_Cryo_char}. We chose an idle time of $40~\mathrm{ns}$ based on this sweep, as the effects of inter-pulse interference are negligible at this point, as indicated in Fig.~\ref{fig_Cryo_char}. In the future, we would like to implement further corrections in pulse shape based on careful analysis of this type of data.

After applying the pulse shape corrections, exchange angle calibrations were performed as described in \cite{Andrews2019}. As our full dynamical decoupling sequences exceed the approximately $10~\mathrm{\mu s}$ waveform memory, we programmed a single unit of the decoupling sequence (i.e., NZ1 or NZM1) into memory and played that waveform in a loop for the given number of repetitions. Since we could not apply the predistortion filter to the entire sequence at once, we applied it to a set of three repetitions of the sequence and used the middle repetition of the now predistorted sequence as the waveform saved in memory.

\section{Mathematical Methods}\label{sec:mathmethods}
\subsection{Noise Description}\label{sec:mathmethods:noise}
\vspace{-0.5em}
\setcounter{figure}{0}

In this analysis, we consider the effects of exchange noise induced by effective gate voltage noise $(\delta J_{ij}(t))$ and hyperfine noise induced by effective magnetic noise $(\delta {\bf B}_{i}(t))$. These noise sources can be approximated as zero-mean Gaussian random variables~\cite{Ladd2012,Eng2015,Cerfontaine2020,Kerckhoff2021}. This allows for a simplification of the noise by converting from time domain to frequency domain using an integrated noise approach~\cite{Reed2016,Kerckhoff2021,Sun2024}. In this approximation, we determine the effect of a given noise source $x(t)$ based on its single-sided power spectral density (PSD), defined as $S_x(f)=2\int_{0}^{\infty} \langle x(0)x(t)\rangle\cos(2\pi f t) dt$. This noise PSD is then convolved with an experiment-specific response function $r(t)$ with corresponding PSD $F(f)$. This response function can either be represented by the envelope of $J_{ij}(t)$ or a filter function formalism~\cite{Burkard2023}. For instance, the PSD of a rectangular window of time $\tau_{\mathrm{p}}$ is given analytically by the $\rect(t)\leftrightarrow \sinc(f)$ Fourier transform pair, effectively filtering out noise above $1/\tau_{\mathrm{p}}$.

\begin{figure}[t!]
	\includegraphics[width=0.75\columnwidth]{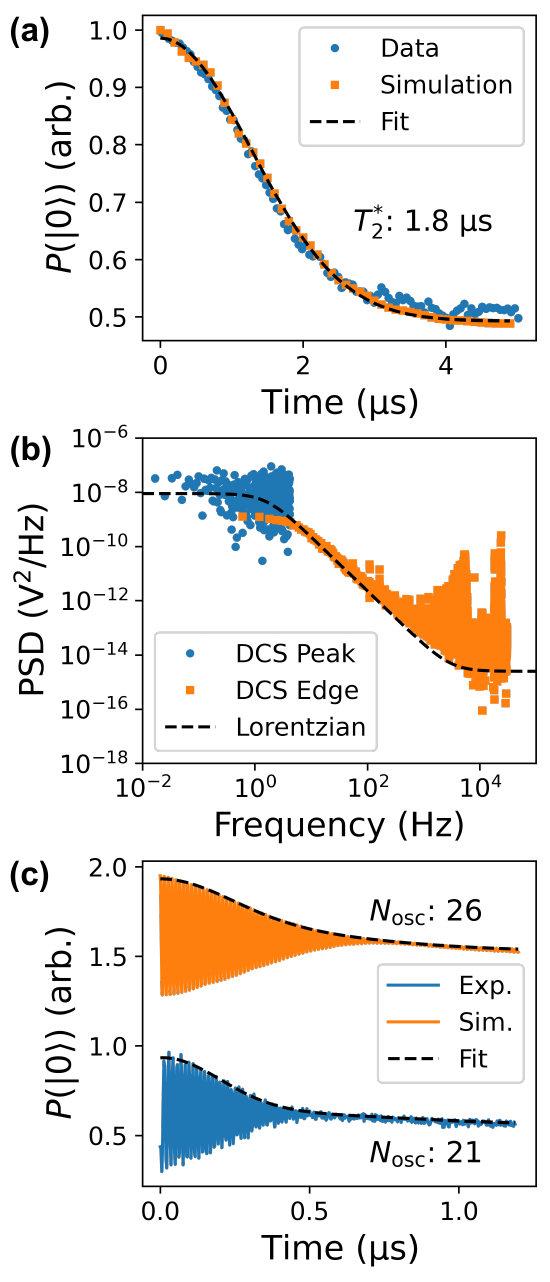}
	\caption{{\bf Noise Environment Extraction:} (a) FID experiment to extract the hyperfine noise PSD. A PSD of the form $S_{B}=B_{1/f}^{2}/f$ is assumed for the fit and subsequent simulation, $B_{1/f}=1~\mathrm{\mu T/\sqrt{Hz}}$. (b) DCS measurement to extract the gate-referred voltage noise PSD by tracking the Coulomb blockade peak and edge. A Lorentzian fits best to the measured voltage noise PSD, $\Delta V=300~\mathrm{\mu V}$, $\tau_{c}=200~\mathrm{ms}$, $A_{w}=40~\mathrm{nV/\sqrt{Hz}}$. (c) $J_{12}=78.1~\mathrm{MHz}$ time-Rabi which couples both exchange and magnetic noise. Gaussian envelopes are used to extract $N_{\mathrm{osc}}$. The experimental data are the same as plotted in Fig.~\ref{fig_1}(c). PSDs from (a,b) are used for the simulation. The simulation data are offset by 1 for visual clarity.}
	\label{fig_A1}
\end{figure}

$J_{ij}(t)$ is modulated by ideal sequential rectangular voltage pulses of width $\tau_{\mathrm{p}}$ with an idle time of $\tau_{\mathrm{i}}$ between pulses. The magnitude of pairwise $J_{ij}(t)$ is determined from an applied gate voltage pulse using a modified two-site Fermi-Hubbard model. Chemical potentials are determined by gate voltage lever arms and the spin state is implied by a singlet-triplet excitation energy~\cite{Pan2020}. In symmetric exchange axis operation, one can condense this complex interaction into a single function by calculating $J_{ij}(V_{tc})$ as a function of tunnel barrier gate voltage. The above allows one to simplify the exchange rotation expression to $\theta_{ij}(\tau)=\frac{J_{ij}(V_{tc})\tau}{\hbar}$. The effect of small perturbations (e.g., voltage noise) from $V_{tc}$ can be approximated linearly as $\delta J_{ij}\approx\delta V_{tc}\frac{\partial J_{ij}}{\partial V}$. In symmetric operation, perturbations to the chemical potential detuning are negligible~\cite{Reed2016}. Additionally, since $J_{ij}$ is exponentially sensitive to $V_{tc}$, exchange noise during idle is also negligible because $J_{ij}(V_{tc})\approx 0$ during these sequence steps~\cite{Pan2020}. In practice, exchange noise is measured as an effective gate voltage noise on the tunnel barrier gate, with the spectrum measured by the DCS used for qubit readout. The measured spectrum for this report is given in Fig.~\ref{fig_A1}(b) and is best fit by a Lorentzian of the form $S_{V}(f)=\frac{\Delta V^{2}\tau_{c}}{2(1+(\pi \tau_{c}f)^{2})}+A_{w}^{2}$, which approximates the PSD of a first-order autoregressive moving average (ARMA) noise model.

${\bf B}=B_{z}$ is a uniform external magnetic field applied to the qubit which defines the Z-projection of the spin system and generates a Zeeman energy splitting between the $m_{z}=\pm 1/2, \pm 3/2$ basis states. The effective spectral character of hyperfine noise depends on how strong the coupling is between the qubit and nearby nuclear spins~\cite{Kerckhoff2021}. In weakly coupled systems, such as isotopically enriched 800 ppm $\mathrm{^{29}Si}$ SiGe heterostructures, the spectrum is typically of the form $S_{B}=B_{1/f}^{2}/f$ within a measured frequency span of $f\in\{10^{-2},10^{6}\}~\mathrm{Hz}$~\cite{Eng2015,Kerckhoff2021}. At higher frequencies, the PSD typically transitions to a $1/f^{2}$ roll-off~\cite{Cerfontaine2020,Kerckhoff2021}. The magnitude of $B_{1/f}$ can be determined from a free induction decay (FID) experiment. From the integrated noise perspective, the qubit decay envelope can be described with the following equation,

\begin{equation}\label{eq:decay}
	C(t)=\exp[-\frac{1}{2}\int_{0}^{\infty} S(f)\tau^{2}\sinc^{2}(\pi f\tau)df],
\end{equation}
\noindent where $\tau$ is the time evolved and $S(f)$ is the noise PSD~\cite{Vezvaee2024}. Eq.~\ref{eq:decay} can be used to either predict the decay envelope from noise PSDs or extract the noise PSDs from a decay envelope~\cite{Eng2015,Reed2016,Vezvaee2024}. For instance, we can predict the expected decay envelope for quasistatic and white noise, described by $S_{qs}(f)=A_{qs}^{2}\delta(f)$ and $S_{w}(f)=A_{w}^{2}$, respectively. The integrations of Eq.~\ref{eq:decay} with $S_{qs}(f)$ and $S_{w}(f)$ correspond to Gaussian and exponential decay envelopes, respectively.

\subsection{Simulation}\label{sec:mathmethods:sim}

DD sequences were simulated via the Dynamiqs Python package~\cite{dynamiqs2025}. Dynamiqs uses the just-in-time compilation package, JAX~\cite{jax2018github}, to perform GPU-accelerated simulations of quantum systems, leveraged here to batch noise trajectories. We use Dynamiqs to simulate EO qubit systems using a stochastic approach where $\delta J_{ij}(t)$ and $\delta {\bf B}_{i}(t)$ are modeled by shot-to-shot uncorrelated zero-mean Gaussian random variables sampled from the frequency spectra outlined in App.~\ref{sec:mathmethods:noise} and averaged over many trials $(N_{\mathrm{avg}}=100)$. We use an HPC cluster housed at HRL Laboratories consisting of eight Nvidia H200 GPUs to execute these simulations. Since $1/f^{\alpha}$ spectra diverge as $f\rightarrow0$, we set a physically motivated low-frequency limit where slower fluctuations are not correlated with measurements (set to $10^{-2}~\mathrm{Hz}$ to capture the duration of a complete DD experiment including averaging, which was approximately $100~\mathrm{s}$). We also set a high-frequency limit of $10^{9}~\mathrm{Hz}$ since the $\sinc$ window from the rectangular exchange pulses sets a $1/f^{2}$ cutoff at $1/\tau_{\mathrm{p}}=10^{8}~\mathrm{Hz}$.

$J_{ij}(t)$ control is performed by sequential rectangular voltage pulses of width $\tau_{\mathrm{p}}=10~\mathrm{ns}$ with an idle time of $\tau_{\mathrm{i}}=40~\mathrm{ns}$ between pulses, reflecting values used in experiment. The magnitude of pairwise $J_{ij}(t)$ from an applied gate voltage is determined using a modified two-site Fermi-Hubbard model following the procedure mentioned in App.~\ref{sec:mathmethods:noise} and described fully in Ref.~\cite{Pan2020}. A uniform in-plane external magnetic field of $50~\mathrm{\mu T}$ is applied in simulation to represent Earth's magnetic field. We consider only the ground orbital three-electron Hilbert space, ignoring excited orbital and valley states.

\vspace{-1em}
\subsection{Drift Extraction}\label{sec:mathmethods:drift}

We fit DD sequence experimental data to an ansatz similar to that used in Refs.~\cite{Andrews2019,Sun2024}. The notable difference is that we add a $\cos$ function to capture the oscillation seen in the NZM1x and NZM1y return fidelities.
\begin{figure}[t!]
	\includegraphics[width=0.95\columnwidth]{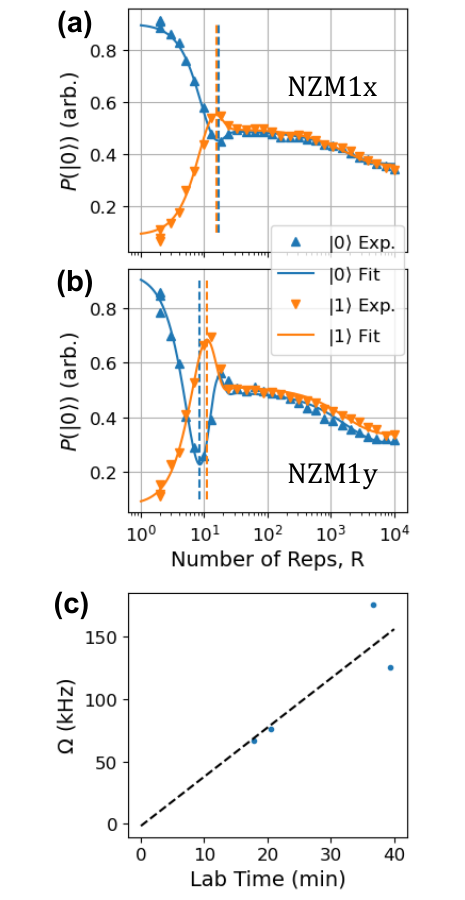}
	\caption{{\bf NZM1x and NZM1y Analysis:} Panels (a) and (b) show the NZM1x and NZM1y data, respectively, along with their corresponding ansatz fits. Dashed vertical lines indicate the oscillation half-period for each ansatz fit, showing a different fitted frequency for each trace. Panel (c) shows the fitted oscillation frequencies from (a) and (b) correlated with the elapsed lab time since the latest calibration. Linear fit parameters are a slope of $65.7~\mathrm{Hz/s}$ and a y-intercept of $-1.7~\mathrm{kHz}$.}
	\label{fig_C1}
\end{figure}
\begin{equation}\label{eq:y_0}
	y_{0}(R)=A+B\cos(\Omega t)(1-p)^{M^{D}}+C(1-q)^{M},
\end{equation}
\begin{equation}\label{eq:y_1}
	y_{1}(R)=A-B\cos(\Omega t)(1-p)^{M^{D}}+C(1-q)^{M}.
\end{equation}

This ansatz is used to extract the return fidelity oscillation frequency, which corresponds to the calibration drift, from the NZM1x and NZM1y data. We then correlate this fitted calibration drift magnitude with the time since the most recent calibration, as given in Fig.~\ref{fig_C1}.

\bibliography{spinQICK_Triangle_bib}

@article{Andrews2019,
  author  = {Andrews, Reed W. and Jones, Cody and Reed, Matthew D. and Jones, Aaron M. and Ha, Sieu D. and Jura, Michael P. and Kerckhoff, Joseph and Levendorf, Mark and Meenehan, Se{\'a}n and Merkel, Seth T. and Smith, Aaron and Sun, Bo and Weinstein, Aaron J. and Rakher, Matthew T. and Ladd, Thaddeus D. and Borselli, Matthew G.},
  title   = {Quantifying error and leakage in an encoded Si/SiGe triple-dot qubit},
  journal = {Nat. Nanotechnol.},
  volume  = {14},
  pages   = {747--750},
  year    = {2019},
  doi     = {10.1038/s41565-019-0500-4}
}

@article{Biercuk2011,
  author  = {Biercuk, M. J. and Doherty, A. C. and Uys, H.},
  title   = {Dynamical decoupling sequence construction as a filter-design problem},
  journal = {J. Phys. B: At. Mol. Opt. Phys.},
  volume  = {44},
  pages   = {154002},
  year    = {2011},
  doi     = {10.1088/0953-4075/44/15/154002}
}

@article{Blumoff2022,
  author  = {Blumoff, Jacob Z. and Pan, Andrew S. and Keating, Tyler E. and Andrews, Reed W. and Barnes, David W. and Brecht, Teresa L. and Croke, Edward T. and Euliss, Larken E. and Fast, Jacob A. and Jackson, Clayton A.C. and Jones, Aaron M. and Kerckhoff, Joseph and Lanza, Robert K. and Raach, Kate and Thomas, Bryan J. and Velunta, Roland and Weinstein, Aaron J. and Ladd, Thaddeus D. and Eng, Kevin and Borselli, Matthew G. and Hunter, Andrew T. and Rakher, Matthew T.},
  title   = {Fast and high-fidelity state preparation and measurement in triple-quantum-dot spin qubits},
  journal = {PRX Quantum},
  volume  = {3},
  pages   = {010352},
  year    = {2022},
  doi     = {10.1103/PRXQuantum.3.010352}
}

@article{Burkard2023,
  author  = {Burkard, Guido and Ladd, Thaddeus D. and Pan, Andrew and Nichol, John M. and Petta, Jason R.},
  title   = {Semiconductor spin qubits},
  journal = {Rev. Mod. Phys.},
  volume  = {95},
  pages   = {025003},
  year    = {2023},
  doi     = {10.1103/RevModPhys.95.025003}
}

@article{Bylander2011,
  author  = {Bylander, Jonas and Gustavsson, Simon and Yan, Fei and Yoshihara, Fumiki and Harrabi, Khalil and Fitch, George and Cory, David G. and Nakamura, Yasunobu and Tsai, Jaw-Shen and Oliver, William D.},
  title   = {Noise spectroscopy through dynamical decoupling with a superconducting flux qubit},
  journal = {Nat. Phys.},
  volume  = {7},
  pages   = {565--570},
  year    = {2011},
  doi     = {10.1038/nphys1994}
}

@article{Cerfontaine2020,
  author  = {Cerfontaine, Pascal and Botzem, Tim and Ritzmann, Julian and Humpohl, Simon Sebastian and Ludwig, Arne and Schuh, Dieter and Bougeard, Dominique and Wieck, Andreas D. and Bluhm, Hendrik},
  title   = {Closed-loop control of a GaAs-based singlet-triplet spin qubit with 99.5{\%} gate fidelity and low leakage},
  journal = {Nat. Commun.},
  volume  = {11},
  pages   = {4144},
  year    = {2020},
  doi     = {10.1038/s41467-020-17865-3}
}

@article{DiVincenzo2000,
  author  = {DiVincenzo, D. P. and Bacon, D. and Kempe, J. and Burkard, G. and Whaley, K. B.},
  title   = {Universal quantum computation with the exchange interaction},
  journal = {Nature},
  volume  = {408},
  pages   = {339--342},
  year    = {2000},
  doi     = {10.1038/35042541}
}

@article{Eng2015,
  author  = {Eng, K. and Ladd, T.D. and Smith, A. and Borselli, M.G. and Kiselev, A.A. and Fong, B.H. and Holabird, K.S. and Hazard, T.M. and Huang, B. and Deelman, P.W. and Milosavljevic, I. and Schmitz, A.E. and Ross, R.S. and Gyure, M.F. and Hunter, A.T.},
  title   = {Isotopically enhanced triple-quantum-dot qubit},
  journal = {Sci. Adv.},
  volume  = {1},
  pages   = {e1500214},
  year    = {2015},
  doi     = {10.1126/sciadv.1500214}
}

@article{Ezzell2023,
  author  = {Ezzell, Nic and Pokharel, Bibek and Tewala, Lina and Quiroz, Gregory and Lidar, Daniel A.},
  title   = {Dynamical decoupling for superconducting qubits: A performance survey},
  journal = {Phys. Rev. Appl.},
  volume  = {20},
  pages   = {064027},
  year    = {2023},
  doi     = {10.1103/PhysRevApplied.20.064027}
}

@misc{FongWadzura2011,
  author        = {Fong, B.H. and Wandzura, S.M.},
  title         = {Universal quantum computation and leakage reduction in the 3-qubit decoherence free subsystem},
  year          = {2011},
  eprint        = {1102.2909},
  archiveprefix = {arXiv},
  url           = {https://arxiv.org/abs/1102.2909}
}

@article{Glattli1997,
  author  = {Glattli, D. C. and Jacques, P. and Kumar, A. and Pari, P. and Saminadayar, L.},
  title   = {A noise detection scheme with 10 mK noise temperature resolution for semiconductor single electron tunneling devices},
  journal = {J. Appl. Phys.},
  volume  = {81},
  pages   = {7350--7356},
  year    = {1997},
  doi     = {10.1063/1.365332}
}

@article{Hahn1950,
  author  = {Hahn, E. L.},
  title   = {Spin echoes},
  journal = {Phys. Rev.},
  volume  = {80},
  pages   = {580--594},
  year    = {1950},
  doi     = {10.1103/PhysRev.80.580}
}

@article{hensgens2017quantum,
  title     = {Quantum simulation of a Fermi--Hubbard model using a semiconductor quantum dot array},
  author    = {Hensgens, Toivo and Fujita, Takafumi and Janssen, Laurens and Li, Xiao and Van Diepen, CJ and Reichl, Christian and Wegscheider, Werner and Das Sarma, Sankar and Vandersypen, Lieven MK},
  journal   = {Nature},
  volume    = {548},
  number    = {7665},
  pages     = {70--73},
  year      = {2017},
  publisher = {Nature Publishing Group UK London},
  doi       = {10.1038/nature23022}
}

@article{Hsieh2012,
  author  = {Hsieh, Chang-Yu and Rene, Alexandre and Hawrylak, Pawel},
  title   = {Herzberg circuit and Berry's phase in chirality-based coded qubit in a triangular triple quantum dot},
  journal = {Phys. Rev. B},
  volume  = {86},
  pages   = {115312},
  year    = {2012},
  doi     = {10.1103/PhysRevB.86.115312}
}

@article{Kerckhoff2021,
  author  = {Kerckhoff, J. and Sun, B. and Fong, B.H. and Jones, C. and Kiselev, A.A. and Barnes, D.W. and Noah, R.S. and Acuna, E. and Akmal, M. and Ha, S.D. and Wright, J.A. and Thomas, B.J. and Jackson, C.A.C. and Edge, L.F. and Eng, K. and Ross, R.S. and Ladd, T.D.},
  title   = {Magnetic gradient fluctuations from quadrupolar ${}^{73}\mathrm{Ge}$ in $\mathrm{Si}$/$\mathrm{Si}\mathrm{Ge}$ exchange-only qubits},
  journal = {PRX Quantum},
  volume  = {2},
  pages   = {010347},
  year    = {2021},
  doi     = {10.1103/PRXQuantum.2.010347}
}

@article{Ladd2012,
  author  = {Ladd, Thaddeus D.},
  title   = {Hyperfine-induced decay in triple quantum dots},
  journal = {Phys. Rev. B},
  volume  = {86},
  pages   = {125408},
  year    = {2012},
  doi     = {10.1103/PhysRevB.86.125408}
}

@article{Meiboom1958,
  author  = {Meiboom, S. and Gill, D.},
  title   = {Modified spin echo method for measuring nuclear relaxation times},
  journal = {Rev. Sci. Instrum.},
  volume  = {29},
  pages   = {688--691},
  year    = {1958},
  doi     = {10.1063/1.1716296}
}

@article{Merkulov2002,
  author  = {Merkulov, I. A. and Efros, Al. L. and Rosen, M.},
  title   = {Electron spin relaxation by nuclei in semiconductor quantum dots},
  journal = {Phys. Rev. B},
  volume  = {65},
  pages   = {205309},
  year    = {2002},
  doi     = {10.1103/PhysRevB.65.205309}
}

@article{Pan2020,
  author  = {Pan, Andrew and Keating, Tyler E. and Gyure, Mark F. and Pritchett, Emily J. and Quinn, Samuel and Ross, Richard S. and Ladd, Thaddeus D. and Kerckhoff, Joseph},
  title   = {Resonant exchange operation in triple-quantum-dot qubits for spin–photon transduction},
  journal = {Quantum Sci. Technol.},
  volume  = {5},
  pages   = {034005},
  year    = {2020},
  doi     = {10.1088/2058-9565/ab86c9}
}

@article{Reed2016,
  author  = {Reed, M. D. and Maune, B. M. and Andrews, R. W. and Borselli, M. G. and Eng, K. and Jura, M. P. and Kiselev, A. A. and Ladd, T. D. and Merkel, S. T. and Milosavljevic, I. and Pritchett, E. J. and Rakher, M. T. and Ross, R. S. and Schmitz, A. E. and Smith, A. and Wright, J. A. and Gyure, M. F. and Hunter, A. T.},
  title   = {Reduced sensitivity to charge noise in semiconductor spin qubits via symmetric operation},
  journal = {Phys. Rev. Lett.},
  volume  = {116},
  pages   = {110402},
  year    = {2016},
  doi     = {10.1103/PhysRevLett.116.110402}
}

@article{Scarola2004,
  author  = {Scarola, V. W. and Park, K. and Sarma, S. Das},
  title   = {Chirality in quantum computation with spin cluster qubits},
  journal = {Phys. Rev. Lett.},
  volume  = {93},
  pages   = {120503},
  year    = {2004},
  doi     = {10.1103/PhysRevLett.93.120503}
}

@article{Sun2024,
  author  = {Sun, Bo and Brecht, Teresa and Fong, Bryan H. and Akmal, Moonmoon and Blumoff, Jacob Z. and Cain, Tyler A. and Carter, Faustin W. and Finestone, Dylan H. and Fireman, Micha N. and Ha, Wonill and Hatke, Anthony T. and Hickey, Ryan M. and Jackson, Clayton A. C. and Jenkins, Ian and Jones, Aaron M. and Pan, Andrew and Ward, Daniel R. and Weinstein, Aaron J. and Whiteley, Samuel J. and Williams, Parker and Borselli, Matthew G. and Rakher, Matthew T. and Ladd, Thaddeus D.},
  title   = {Full-permutation dynamical decoupling in triple-quantum-dot spin qubits},
  journal = {PRX Quantum},
  volume  = {5},
  pages   = {020356},
  year    = {2024},
  doi     = {10.1103/PRXQuantum.5.020356}
}

@article{Taylor2007,
  author  = {Taylor, J. M. and Petta, J. R. and Johnson, A. C. and Yacoby, A. and Marcus, C. M. and Lukin, M. D.},
  title   = {Relaxation, dephasing, and quantum control of electron spins in double quantum dots},
  journal = {Phys. Rev. B},
  volume  = {76},
  pages   = {035315},
  year    = {2007},
  doi     = {10.1103/PhysRevB.76.035315}
}

@article{Weinstein2005,
  author  = {Weinstein, Yaakov S. and Hellberg, C. Stephen},
  title   = {Energetic suppression of decoherence in exchange-only quantum computation},
  journal = {Phys. Rev. A},
  volume  = {72},
  pages   = {022319},
  year    = {2005},
  doi     = {10.1103/PhysRevA.72.022319}
}

@article{Weinstein2023,
  author  = {Weinstein, Aaron J. and Reed, Matthew D. and Jones, Aaron M. and Andrews, Reed W. and Barnes, David and Blumoff, Jacob Z. and Euliss, Larken E. and Eng, Kevin and Fong, Bryan H. and Ha, Sieu D. and Hulbert, Daniel R. and Jackson, Clayton A. C. and Jura, Michael and Keating, Tyler E. and Kerckhoff, Joseph and Kiselev, Andrey A. and Matten, Justine and Sabbir, Golam and Smith, Aaron and Wright, Jeffrey and Rakher, Matthew T. and Ladd, Thaddeus D. and Borselli, Matthew G.},
  title   = {Universal logic with encoded spin qubits in silicon},
  journal = {Nature},
  volume  = {615},
  pages   = {817--822},
  year    = {2023},
  doi     = {10.1038/s41586-023-05777-3}
}

@article{West2012,
  author  = {West, J. R. and Fong, B. H.},
  title   = {Exchange-only dynamical decoupling in the three-qubit decoherence free subsystem},
  journal = {New J. Phys.},
  volume  = {14},
  pages   = {083002},
  year    = {2012},
  doi     = {10.1088/1367-2630/14/8/083002}
}

@misc{spinQICK2025,
  author  = {Wessels, Abigail L. and Oriani, Andrew E.},
  title   = {spinQICK},
  year    = {2025},
  journal = {GitHub repository},
  url     = {https://github.com/HRL-Laboratories/spinqick#}
}

@article{Ding2024,
  author  = {Ding, Chunyang and Di Federico, Martin and Hatridge, Michael and Houck, Andrew and Leger, Sebastien and Martinez, Jeronimo and Miao, Connie and Schuster, David I. and Stefanazzi, Leandro and Stoughton, Chris and Sussman, Sara and Treptow, Ken and Uemura, Sho and Wilcer, Neal and Zhang, Helin and Zhou, Chao and Cancelo, Gustavo},
  title   = {Experimental advances with the QICK (Quantum Instrumentation Control Kit) for superconducting quantum hardware},
  journal = {Phys. Rev. Res.},
  volume  = {6},
  pages   = {013305},
  year    = {2024},
  doi     = {10.1103/PhysRevResearch.6.013305}
}

@article{Acuna2024,
  author  = {Acuna, Edwin and Broz, Joseph D. and Shyamsundar, Kaushal and Mei, Antonio B. and Feeney, Colin P. and Smetanka, Valerie and Davis, Tiffany and Lee, Kangmu and Choi, Maxwell D. and Boyd, Brydon and Suh, June and Ha, Wonill and Jennings, Cameron and Pan, Andrew S. and Sanchez, Daniel S. and Reed, Matthew D. and Petta, Jason R.},
  title   = {Coherent control of a triangular exchange-only spin qubit},
  journal = {Phys. Rev. Appl.},
  volume  = {22},
  pages   = {044057},
  year    = {2024},
  doi     = {10.1103/PhysRevApplied.22.044057}
}

@misc{dynamiqs2025,
  author = {Pierre Guilmin and Adrien Bocquet and {\'{E}}lie Genois and Daniel Weiss and Ronan Gautier},
  title  = {Dynamiqs: an open-source Python library for GPU-accelerated and differentiable simulation of quantum systems},
  year   = {2025},
  url    = {https://github.com/dynamiqs/dynamiqs}
}

@article{Vezvaee2024,
  author  = {Vezvaee, Arian and Shitara, Nanako and Sun, Shuo and Montoya-Castillo, Andrés},
  title   = {Fourier transform noise spectroscopy},
  journal = {npj Quantum Inf.},
  volume  = {10},
  pages   = {52},
  year    = {2024},
  doi     = {10.1038/s41534-024-00841-w}
}

@article{Madzik2025,
  author  = {M{a}dzik, Mateusz T. and Luthi, Florian and Guerreschi, Gian Giacomo and Mohiyaddin, Fahd A. and Borjans, Felix and Chadwick, Jason D. and Curry, Matthew J. and Ziegler, Joshua and Atanasov, Sarah and Bavdaz, Peter L. and Connors, Elliot J. and Corrigan, J. and Ercan, H. Ekmel and Flory, Robert and George, Hubert C. and Harpt, Benjamin and Henry, Eric and Islam, Mohammad M. and Khammassi, Nader and Keith, Daniel and Lampert, Lester F. and Mladenov, Todor M. and Morris, Randy W. and Nethwewala, Aditi and Neyens, Samuel and Otten, Ren{\'e} and Osuna Ibarra, Linda P. and Patra, Bishnu and Pillarisetty, Ravi and Premaratne, Shavindra and Ramsey, Mick and Risinger, Andrew and Rooney, John D. and Savytskyy, Rostyslav and Watson, Thomas F. and Zietz, Otto K. and Matsuura, Anne Y. and Pellerano, Stefano and Bishop, Nathaniel C. and Roberts, Jeanette and Clarke, James S.},
  title   = {Operating two exchange-only qubits in parallel},
  journal = {Nature},
  volume  = {647},
  pages   = {870--875},
  year    = {2025},
  doi     = {10.1038/s41586-025-09767-5}
}

@article{Kempe2000,
  author  = {Bacon, D. and Kempe, J. and Lidar, D. A. and Whaley, K. B.},
  title   = {Universal fault-tolerant quantum computation on decoherence-free subspaces},
  journal = {Phys. Rev. Lett.},
  volume  = {85},
  pages   = {1758--1761},
  year    = {2000},
  doi     = {10.1103/PhysRevLett.85.1758}
}

@article{Stefanazzi2022,
  author  = {Stefanazzi, Leandro and Treptow, Kenneth and Wilcer, Neal and Stoughton, Chris and Bradford, Collin and Uemura, Sho and Zorzetti, Silvia and Montella, Salvatore and Cancelo, Gustavo and Sussman, Sara and Houck, Andrew and Saxena, Shefali and Arnaldi, Horacio and Agrawal, Ankur and Zhang, Helin and Ding, Chunyang and Schuster, David I.},
  title   = {The QICK (Quantum Instrumentation Control Kit): Readout and control for qubits and detectors},
  journal = {Rev. Sci. Instrum.},
  volume  = {93},
  pages   = {044709},
  year    = {2022},
  doi     = {10.1063/5.0076249}
}

@misc{sadi2026,
  author        = {Mohammad Abdullah Sadi and Tiamike Dudley and Luca Basso and Thomas Poirier and James H. Edgar and Jacob Henshaw and Peter A. Bermel and Yong P. Chen and Andrew Mounce},
  title         = {Landau Zener interaction enhanced quantum sensing in spin defects of hexagonal boron nitride},
  year          = {2026},
  eprint        = {2601.05013},
  archiveprefix = {arXiv},
  url           = {https://arxiv.org/abs/2601.05013}
}

@misc{koong2025,
  author        = {Zhe Xian Koong and Urs Haeusler and Jan M. Kaspari and Christian Schimpf and Benyam Dejen and Ahmed M. Hassanen and Daniel Graham and Ailton J. Garcia Jr. and Melina Peter and Edmund Clarke and Maxime Hugues and Armando Rastelli and Doris E. Reiter and Mete Atatüre and Dorian A. Gangloff},
  title         = {Coherent control of quantum-dot spins with cyclic optical transitions},
  year          = {2025},
  eprint        = {2509.14445},
  archiveprefix = {arXiv},
  url           = {https://arxiv.org/abs/2509.14445}
}

@misc{vora2024,
  author        = {Neel R. Vora and Yilun Xu and Akel Hashim and Neelay Fruitwala and Ho Nam Nguyen and Haoran Liao and Jan Balewski and Abhi Rajagopala and Kasra Nowrouzi and Qing Ji and K. Birgitta Whaley and Irfan Siddiqi and Phuc Nguyen and Gang Huang},
  title         = {ML-powered FPGA-based real-time quantum state discrimination enabling mid-circuit measurements},
  year          = {2024},
  eprint        = {2406.18807},
  archiveprefix = {arXiv},
  url           = {https://arxiv.org/abs/2406.18807}
}

@article{broz2025,
  author  = {Broz, Joseph D. and Hoke, Jesse C. and Acuna, Edwin and Petta, Jason R.},
  title   = {Demonstration of an always-on exchange-only spin qubit},
  journal = {Nat. Commun.},
  volume  = {17},
  pages   = {4794},
  year    = {2026},
  doi     = {10.1038/s41467-026-70943-w}
}

@article{Chadwick2025,
  author  = {Chadwick, Jason D. and Guerreschi, Gian Giacomo and Luthi, Florian and Madzik, Mateusz T. and Mohiyaddin, Fahd A. and Prabhu, Prithviraj and Schmitz, Albert T. and Litteken, Andrew and Premaratne, Shavindra and Bishop, Nathaniel C. and Matsuura, Anne Y. and Clarke, James S.},
  title   = {Short two-qubit pulse sequences for exchange-only spin qubits in two-dimensional layouts},
  journal = {Phys. Rev. A},
  volume  = {111},
  pages   = {052616},
  year    = {2025},
  doi     = {10.1103/PhysRevA.111.052616}
}

@software{jax2018github,
  author  = {James Bradbury and Roy Frostig and Peter Hawkins and Matthew James Johnson and Chris Leary and Dougal Maclaurin and George Necula and Adam Paszke and Jake Vander{P}las and Skye Wanderman-{M}ilne and Qiao Zhang},
  title   = {{JAX}: composable transformations of {P}ython+{N}um{P}y programs},
  url     = {http://github.com/jax-ml/jax},
  version = {0.3.13},
  year    = {2018}
}

@article{Guo2024,
  author  = {Guo, Liang-Liang and Duan, Peng and Zhang, Sheng and Yang, Xin-Xin and Zhang, Chi and Du, Lei and Zhang, Hai-Feng and Tao, Hao-Ran and Wang, Tian-Le and Jia, Zhi-Long and Chen, Zhao-Yun and Guo, Guo-Ping},
  title   = {Universal scalable characterization and correction of pulse distortions in controlled quantum systems},
  journal = {Phys. Rev. Appl.},
  volume  = {21},
  pages   = {064060},
  year    = {2024},
  doi     = {10.1103/PhysRevApplied.21.064060}
}

@article{hrl2026digitally,
  author  = {{\relax Members of the HRL Quantum Team and Collaborators}},
  title   = {A digitally controlled silicon quantum processing unit},
  journal = {Nature},
  volume  = {655},
  pages   = {1154--1159},
  year    = {2026},
  doi     = {10.1038/s41586-026-10754-7}
}

@article{broz2026leakage,
  author  = {Broz, Joseph D. and Hoke, Jesse C. and Acuna, Edwin and Petta, Jason R.},
  title   = {Leakage-protected idle operation of a triangular exchange-only spin qubit},
  journal = {PRX Quantum},
  volume  = {7},
  pages   = {020324},
  year    = {2026},
  url     = {https://doi.org/10.1103/sljc-cmdw}
}

@misc{xu2023,
  author        = {Xu, Yilun and Huang, Gang and Fruitwala, Neelay and Rajagopala, Abhi and Naik, Ravi K. and Nowrouzi, Kasra and Santiago, David I. and Siddiqi, Irfan},
  title         = {QubiC 2.0: An extensible open-source qubit control system capable of mid-circuit measurement and feed-forward},
  year          = {2023},
  eprint        = {2309.10333},
  archiveprefix = {arXiv},
  url           = {https://arxiv.org/abs/2309.10333}
}

\end{document}